\documentclass[aps,prb,preprint,unsortedaddress,showpacs]{revtex4}

\newcommand{\be}{\begin{equation}}
\newcommand{\ee}{\end{equation}}

\usepackage{graphicx}
\usepackage{amsmath}
\graphicspath{{.}{./EPS/}}

\begin{document}

\title{Two-dimensional hole precession
in an all-semiconductor spin field effect transistor}

\author{Marco G.~Pala}
\affiliation{Institut f\"ur Theoretische Festk\"orperphysik,
Universit\"at Karlsruhe, D-76128 Karlsruhe, Germany}
\affiliation{Dipartimento di Ingegneria dell'Informazione, via
Diotisalvi 2, I-56126 Pisa, Italy}

\author{Michele Governale}
\affiliation{Institut f\"ur Theoretische Festk\"orperphysik,
Universit\"at Karlsruhe, D-76128 Karlsruhe, Germany}
\affiliation{NEST-INFM \& Scuola Normale Superiore, piazza dei Cavalieri 7, 
I-56100 Pisa, Italy}

\author{J\"urgen K\"onig}
\affiliation{Institut f\"ur Theoretische Festk\"orperphysik,
Universit\"at Karlsruhe, D-76128 Karlsruhe, Germany}
\affiliation{Institut f\"ur Theoretische Physik III,  
Ruhr-Universit\"at Bochum, D-44780 Bochum, Germany}

\author{Ulrich Z\"ulicke}
\affiliation{Institut f\"ur Theoretische Festk\"orperphysik,
Universit\"at Karlsruhe, D-76128 Karlsruhe, Germany}
\affiliation{Institute of Fundamental Sciences, Massey 
University, Private Bag 11~222, Palmerston North, New Zealand}
\thanks{Present and permanent address.}

\author{Giuseppe Iannaccone}
\affiliation{Dipartimento di Ingegneria dell'Informazione, via
Diotisalvi 2, I-56126 Pisa, Italy}

\date{\today}

\begin{abstract}

We present a theoretical study of a spin field-effect transistor
realized in a quantum well formed in a p--doped
ferromagnetic-semiconductor-
nonmagnetic-semiconductor-ferromagnetic-semiconductor
hybrid structure. Based on an envelope-function approach for the
hole bands in the various regions of the transistor, we derive the
complete theory of coherent transport through the device, which
includes both heavy- and light-hole subbands, proper modeling of
the mode matching at interfaces, integration over injection angles,
Rashba spin precession, interference effects due to multiple
reflections, and gate-voltage dependences. Numerical results for
the device current as a function of externally tunable parameters
are in excellent agreement with approximate analytical formulae.

\end{abstract}

\pacs{85.75.Hh, 72.25.-b, 73.23.Ad}

\maketitle

\section{INTRODUCTION}

Spintronics has attracted great interest in the scientific
community,\cite{wolf,lossbook} advocating the use of the spin degree of
freedom in electronic devices. Combining this idea with mesoscopic
transport has stimulated investigations of coherent spin-dependent
phenomena. Many proposed device setups exploit the effect of
spin-orbit coupling on the carrier motion.
\cite{devices1,devices2,devices3,devices4,devices5,devices6,devices7} 
The most popular proposal for a coherent spintronic device 
is the spin field-effect
transistor (spin FET) proposed by Datta and Das.\cite{spinfet1,spinfet2}
It consists of a two-dimensional (2D) electron gas confined in a
semiconductor heterostructure that is attached to two ferromagnetic
contacts acting as source and drain. Majority--spin electrons
injected from the source experience a spin precession due to the
Rashba effect\cite{rashba,byra,lommer} if the magnetization
direction in the source contact is parallel to the direction of
current flow or perpendicular to the plane of the 2D electron gas.
Tunability of the spin-orbit coupling strength by gate voltages
enables external control of this spin precession and, hence,
manipulation of the current transmitted at the second ferromagnetic
contact. Besides gate--voltage control of the Rashba spin-orbit
coupling experienced by 2D electrons, which has been successfully
demonstrated experimentally,\cite{sdhexpts1,sdhexpts2,sdhexpts3} 
efficient injection of
spin--polarized electrons from ferromagnetic contacts into the
nonmagnetic part of the spin FET is a key ingredient for device
operation. The obvious challenges involved in the fabrication
of hybrid systems consisting of metallic and semiconducting parts,
as well as a physical limitation\cite{schmidt} to the amount of
spin injection that can be achieved in the absence of tunnel
barriers at the interfaces, have so far prevented the realization
of any spin FET device. A possible solution to circumvent these
difficulties may be provided by the use of diluted magnetic
semiconductors\cite{ohno} as source and drain. This motivates our
present study where we investigate transport through 2D hybrid
structures with ferromagnetic contacts realized in semiconductor
heterostructures. An important aspect of our work deals with the
fact that ferromagnetic (III,Mn)V compounds are intrinsically
p-doped, implying that currents are carried by holes rather than
electrons. The spin properties of carrier states in the
intrinsically p-like valence bands of III-V semiconductors are
very different from that in the s-like conduction band. To begin
with, several valence bands with different effective masses exist.
More importantly, however, spin-orbit coupling in the valence bands
has a more complicated structure than that of conduction--band
electrons.\cite{winkler1,winkler2}

\begin{figure}
\includegraphics[width=8cm]{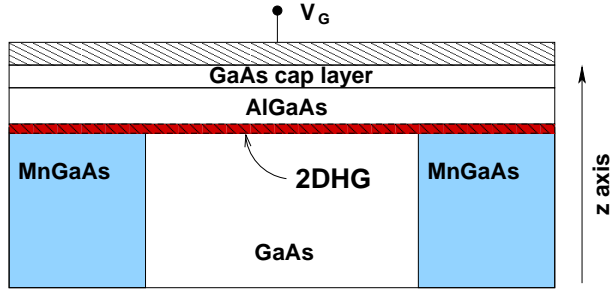}
\caption{Schematic illustration of the proposed device. The
two-dimensional hole gas (2DHG) in the GaAs part is attached to
spin-polarized source and drain contacts, formed by 2DHGs in the
MnGaAs parts. The gate electrode on the top controls both the
carrier concentration and the Rashba spin-orbit coupling strength.}
\label{f1}
\end{figure}

The aim of this article is a detailed study of an all-semiconductor
spin field-effect transistor in which the conducting channel is
provided by a 2D hole gas (2DHG). The device structure we propose
is depicted in Fig.~\ref{f1}. A MnGaAs/GaAs/MnGaAs heterostructure
is overgrown\cite{ceorefs1,ceorefs2} in the $z$ direction with AlGaAs such
that a 2DHG forms at the interface. In fact, Mn doping is only
required within the quantum well formed at the interface to the
AlGaAs layer, but the Mn ions outside the well do not disturb. In
the proposed setup, source and drain are defined by 2D quantum
wells accommodating spin polarized holes. The carriers in the
entire 2DHG are subject to the Rashba effect which leads to spin
precession. The strength of the Rashba spin-orbit coupling can be
tuned by a gate voltage applied to the top of the
sample.\cite{tunekso}

The key ingredient for the functionality of the device is the
tunable Rashba spin precession in the 2DHG. In
Ref.~\onlinecite{pala}, we studied this part within a simplified
model allowing for an analytic treatment, which enabled us to
discuss features that are universal for both electron and hole
transport. In the present paper, we aim at a more complete
numerical treatment of transport through the entire device. This
includes describing the semiconductor valence bands by a $4 \times
4$ Kohn-Luttiger Hamiltonian instead of restricting ourselves to
heavy-hole bands only. We take into account the matching properties
of the modes at the interfaces to the source and drain contacts,
which automatically includes interference effects due to multiple
reflection. As in Ref.~\onlinecite{pala}, we allow for all possible
injection angles instead of restricting to a quasi-one-dimensional setup.
In our analysis we study the purely ballistic regime.
Scattering due to impurities or to the lateral
finite size of the device due to the functional form of the spin-orbit
coupling Hamiltonian for holes affects the total angular
momentum $\mathbf{j}$ and not only the spin.
This leads to extra damping of spin-precession and should
be minimized to improve device functionality.
Finally, for addressing the response of transport to the
gate voltage, we employ a capacitive model that takes the variation
of both the carrier density and the Rashba coupling strength into
account.

The article is organized as follows. We introduce, in
Sec.~\ref{s1}, the Kohn-Luttinger Hamiltonian for the ferromagnetic
and nonmagnetic parts of the semiconductor quantum well. In the
following Section~\ref{s2}, we describe the mode-matching technique
used to calculate transmission coefficients for transport through
the structure. Our results from numerical simulations of transport
are presented in Section~\ref{s:numsim}. After discussing
the case of only one interface between a ferromagnetic and a
nonmagnetic 2DHG, we turn our attention 
to the full transistor geometry with two
ferromagnetic contacts. In the latter case, interference effects
appear due to multiple reflections at the interfaces. In
Section~\ref{s3}, we describe an analytical model that approximates
the numerical simulations very well and helps us to understand how
the precession length $L_{\text {so}}$ depends on the Fermi energy
$E_{\text F}$ in the 2DHG. In Section~\ref{s4}, we address the
response to external gate voltages. The possibility to control
the spin precession by a single gate voltage $V_{\text G}$ that
simultaneously modifies Fermi energy and Rashba spin-orbit coupling
is discussed in detail.

\section{Envelope-function description of 2D valence-band states}
\label{s1}

In this section, we obtain effective Hamiltonians that describe 
the valence bands in the different regions of the spin FET,
namely, the ferromagnetic source and drain contacts doped with
Mn$^{2+}$ ions and the undoped nonmagnetic channel in between.
In all these regions, holes are confined within a 2D quantum well. 
We use an envelope-function description\cite{chow,bastard,fernandez} 
of the 2D system. The Hamiltonian for the nonmagnetic semiconductor
$H_{\text{p}}$ is the sum of a 2D quantum-well Hamiltonian
$H_{\text{2D}}$ plus a Rashba term $H_{\text{rs}}$, which arises
due to the asymmetry of the confinement potential $V_{\text{con}}
(z)$. On the other hand, the total Hamiltonian for the
ferromagnetic contacts $H_{\text{f}}$ is given by $H_{\text{p}}$
plus the term $H_{\text{pd}}$, which takes into account the
coupling between the p-like valence holes and the half-filled
d-shell Mn$^{2+}$ ions with spin $S=5/2$. 

Bulk systems host heavy- and light-hole bands with total angular
momentum $j=3/2$. These bands are degenerate at the band edge and
are well separated, due to spin-orbit coupling, from the split-off
bands with total angular momentum $j=1/2$. In quantum wells, each
of these bands is transformed in a sequence of quasi--2D subbands,
and the degeneracy between heavy and light-hole bands is lifted.
In the following, we are interested in the situation at low
carrier concentrations such that only the lowest subbands, one
heavy--hole band (HH1) and one light-hole band (LH1), have to be
taken into account. 
This is the simplest realistic case that occurs when 
the triangular quantum well is narrow enough
to sufficiently lift the energy of higher subbands.
Inclusion of higher energy subbands, in particular HH2,
would be straightforward, but would only lead to higher
order corrections. Indeed, the low hole density 
typically present in experiments implies that
the only occupied propagating modes are in HH1,
whose shape is influenced by the band mixing\cite{chow} with LH1 and
in a negligible way with HH2. Coupling with HH2 would
affect the shape of LH1, and therefore only evanescent modes.
In the basis of total angular momentum,
\be
\label{basis}
\begin{array}{l}
|1\rangle= |j=3/2,j_z=3/2\rangle\\
|2\rangle= |j=3/2,j_z=-1/2\rangle\\
|3\rangle= |j=3/2,j_z=1/2\rangle\\
|4\rangle= |j=3/2,j_z=-3/2\rangle
\end{array} \; ,
\ee
the Hamiltonian $H_{\text{2D}}$ reads\cite{chow}
\be
\label{h2d}
H_{\text{2D}}=\left( \begin{array}{cccc}
hh & d & 0 & 0\\ d^* & lh & 0 & 0\\
0 & 0 & lh & d\\ 0 & 0 & d^* & hh \end{array} \right) \;,
\ee
with 
\be
\label{h2dterms}
\begin{array}{l}
hh= E^{\text{hh1}} + \frac{\hbar^2}{2 m_{\text{hh} \parallel}} 
k^2\\ 
lh= E^{\text{lh1}} + \frac{\hbar^2}{2 m_{\text{lh} \parallel}} 
k^2\\
d=-\frac{\sqrt{3} \hbar^2}{2m} \, \overline{\gamma} \, N \, 
k^2 \, e^{-i 2\alpha} \; .
\end{array}
\ee
Here we have adopted the momentum-space representation in polar
coordinates for the wave vector ${\bf k}_\parallel= (k \cos \alpha,
k \sin\alpha )$ in the 2D plane. The quantities $E^{\text{hh1}}$
and $E^{\text{lh1}}$ in Eq.~(\ref{h2dterms}) are the subband-bottom
energies deriving from the solution of the Schr\"odinger problem
for the triangular well, 
and the factor $N$ takes into account the scalar product
between the envelope functions $f^{\text{hh1,lh1}}$  for the HH1
and LH1 subbands. We observe that in Eq.~(\ref{h2d}) there is no
coupling between LH and HH subbands with the same sign of $j_z$
since the corresponding matrix element is proportional to the
vanishing integral $\langle f^{\text{hh1}} |k_z| f^{\text{lh1}}
\rangle$.

\begin{figure}
\includegraphics[width=8cm]{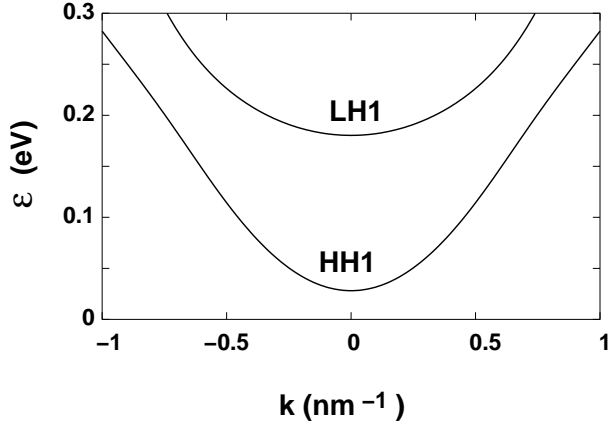}
\caption{Dispersion of the lowest quasi--2D subbands (HH1 and LH1)
of a semiconductor quantum well when no exchange field and no
spin-orbit coupling are present. The subband-bottom
energies are computed using $e E_z$ equal to $4 \times 10^{7}$ eV/m
and vertical masses of $m_{\text{hh} z}=0.38 m_0$ and  
$m_{\text{lh} z}=0.09 m_0$.}
\label{f2}
\end{figure}

The effective masses appearing in Eq.~(\ref{h2dterms}) are given by
\be
\begin{array}{l}
m_{\text{hh} \parallel}=m/(\gamma_1 + \overline{\gamma})\\
m_{\text{lh} \parallel}=m/(\gamma_1 - \overline{\gamma})
\end{array} \; , 
\ee 
where the two coefficients $\gamma_1$ and $ \overline{\gamma}=(
\gamma_2+\gamma_3)/2$ are the Luttinger parameters,\cite{luttinger}
taken in the so-called axial approximation.\cite{altarelli} In
Fig.~\ref{f2} we show the HH1 and the LH1 subband dispersion
relations as a function of the magnitude $k$ of 2D wave vector.

\begin{figure}[b]
\includegraphics[width=8cm]{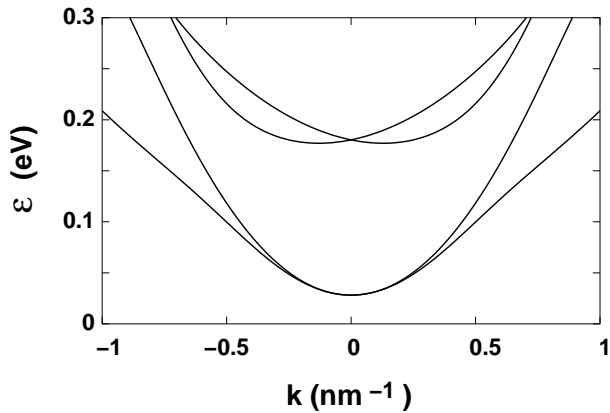}
\caption{The first subbands (HH1 and LH1) of a semiconductor
quantum well in the presence of Rashba spin-orbit coupling, which
removes the spin degeneracy. The coupling constant $\langle \beta
E_z\rangle$ is 0.15 eV nm.}
\label{f3}
\end{figure}

\begin{figure}[t]
\includegraphics[width=8cm]{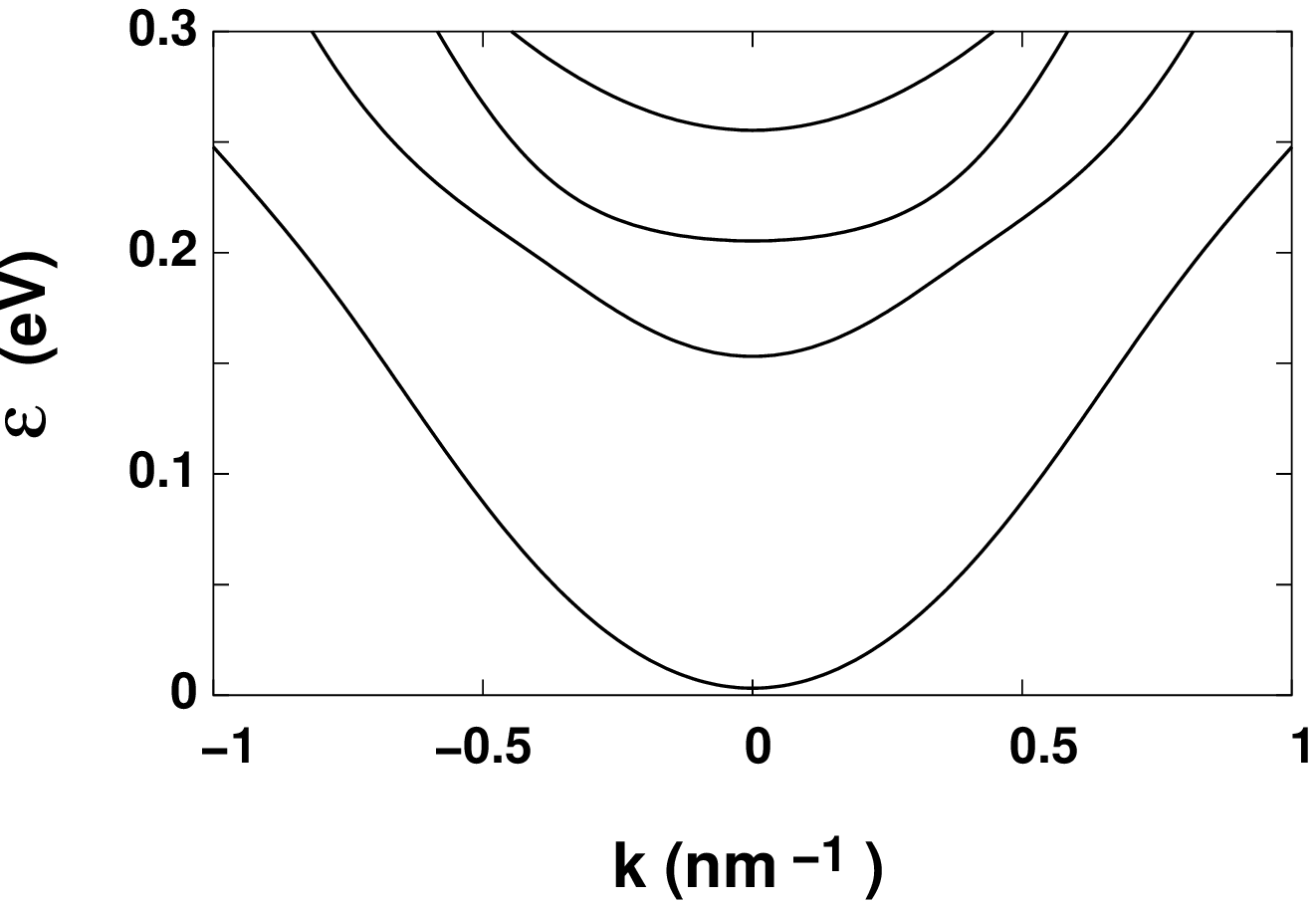}
\caption{The ferromagnetic-semiconductor 2D quantum-well subbands
for the case of magnetization perpendicular to the 2DHG plane. The
interaction constant is $J_{\text{pd}} = 0.06$~eV~nm$^3$, the
Manganese concentration is $N_{\text{Mn}}=1$~nm$^{-3}$, and
$\langle\beta E_z\rangle = 0.15$~eV~nm.}
\label{f4}
\end{figure}   

The Rashba spin-orbit coupling arises from the structural inversion
asymmetry\cite{rashba} due to an asymmetric confining potential. It
is described by the Hamiltonian $H_{\text{rs}}=\beta ({\bf k}
\times {\bf E}) \cdot {\bf j}$. Here, ${\bf E}$ is the electric
field due to the confining potential $V_{\text{con}} (z)$, and
$\beta$ is a material parameter.\cite{winkler1,winkler2} 
In our system, ${\bf E}=E_{z} \, \hat{{\bf z}}$, and $H_{\text{rs}}$ 
reads
\be
\label{hrs}
H_{\text{rs}}=i \beta E_{z} k
\left( \begin{array}{cccc}
0 & 0 & \frac{\sqrt{3}}{2} e^{-i \alpha} & 0\\
0 & 0 & -e^{i \alpha} & \frac{\sqrt{3}}{2} e^{-i \alpha} \\
-\frac{\sqrt{3}}{2} e^{i \alpha} & e^{-i \alpha} & 0 & 0\\
0 & -\frac{\sqrt{3}}{2} e^{i \alpha} & 0 & 0
\end{array} \right).
\ee
In Fig.~\ref{f3} we show the HH and the LH subbands for the 
nonmagnetic semiconductor  as a function of $k$, obtained by
diagonalization of $H_{\text{p}} = H_{\text{2D}} + H_{\text{rs}}$.
We see that the splitting at small wave vectors is linear for
the light-hole subbands but cubic for the heavy-hole subbands.

The Hamiltonian for the ferromagnetic-semiconductor part is given
by $H_{\text{f}}=H_{\text{2D}}+H_{\text{rs}}+H_{\text{pd}}$. We use
a phenomenological description of the ferromagnetic semiconductor,
in which local moments with $S=5/2$ from Mn$^{2+}$ ions are
antiferromagnetically coupled to the itinerant holes.
\cite{models1,models2,models3}
In a mean-field treatment, combined with a virtual-crystal 
approximation,\cite{mf1,mf2,mf3,mf4} itinerant holes experience an exchange
field ${\bf h}=J_{\text{pd}} N_{\text{Mn}}\langle {\bf S} \rangle$,
where the average Mn-ion spin ${\bf S}$ has direction $\hat{{\bf n}}=
(\cos{\phi} \sin{\theta},\sin{\phi}\sin{\theta},\cos{\theta})$,
$N_{\text{Mn}}$ is the doping concentration, and $J_{\text{pd}}$
describes the coupling strength. This exchange-coupling field is
accounted for in $H_{\text{pd}}= {\bf h}\cdot {\bf \sigma}$, where
the spin matrices ${\bf \sigma}$ in the basis given in
Eq.~(\ref{basis}) are
\be
\label{sigmax}
\sigma_x =
\left( \begin{array}{cccc}
0 & 0 &\frac{1}{2\sqrt{3}} & 0\\ 
0 & 0 &\frac{1}{3} &\frac{1}{2\sqrt{3}}\\ 
\frac{1}{2\sqrt{3}} & \frac{1}{3} & 0 & 0\\
0 & \frac{1}{2\sqrt{3}} & 0 & 0
\end{array} \right) \; ,
\ee
\be
\label{sigmay}
\sigma_y = i
\left( \begin{array}{cccc}
0 & 0 &-\frac{1}{2\sqrt{3}} & 0\\ 
0 & 0 &\frac{1}{3} &-\frac{1}{2\sqrt{3}}\\ 
\frac{1}{2\sqrt{3}} & -\frac{1}{3} & 0 & 0\\
0 & \frac{1}{2\sqrt{3}} & 0 & 0
\end{array} \right) \; ,
\ee
and
\be
\label{sigmaz}
\sigma_z =
\left( \begin{array}{cccc}
\frac{1}{2} & 0 & 0 & 0\\ 
0 &-\frac{1}{6}  & 0 & 0\\ 
0 & 0 & \frac{1}{6} & 0\\
0 & 0 & 0 & -\frac{1}{2}
\end{array} \right) \; .
\ee
In Fig.~\ref{f4} we show the HH and the LH subbands in the
ferromagnetic-semiconductor contacts as a function of $k$, obtained
by diagonalization of $H_{\text{f}}$. The splitting of the two
heavy-hole subbands for small values of $k$ leads to full
polarization at low densities. The magnetization direction in the
Figure is perpendicular to the 2DHG plane, $\hat{{\bf n}}=(0,0,1)$.

\section{Quantum states for holes propagating through the spin FET}
\label{s2}

We calculate coherent transport through the spin FET using the
scattering formalism described, e.g., in Ref.~\onlinecite{datta}.
It relates the current to transmission amplitudes for scattering
states defined in the contacts. To obtain these, proper matching
of wave functions at interfaces is required, that we describe in
this section. The imposed conditions at the interface are the
continuity of the wave function and conservation of the component
of probability current that is perpendicular to the interface. To
illustrate the subtleties associated with the second condition, let
us consider an interface between a ferromagnetic and a nonmagnetic
region at $x=x_{0}$. The continuity equation for the current is
\be
\label{condition}
v_{x}^{\text{f}} \psi^{\text{f}}(x_0,y)=
v_{x}^{\text{p}} \psi^{\text{p}}(x_0,y), 
\ee
with 
\be
\label{velocity}
v_{x}^{\text{f},\text{p}}=\frac{1}{\hbar}\frac{\partial 
H_{\text{f},\text{p}}}{\partial k_{x}} \, ,
\ee
which derives from the operator relation $\hat{v}_x=\frac{i}{\hbar}
[\hat{H},\hat{x}]$. Note that, due to the presence of spin-orbit
coupling, the derivative of wave functions needs not to be
continuous at $x=x_0$. Instead, Eq.~(\ref{condition}) guarantees
current conservation.

In both the ferromagnetic and nonmagnetic regions, four different
channels are available, $i=1,\ldots,4$, associated with the
four-dimensional Hilbert space of the valence-band subspace under
consideration. Let us consider a wave incoming from the
ferromagnetic region with wave vector ${\bf k}^{\text{I}}_i=
k^{\text{I}}_i \, (\cos{\alpha},\sin{\alpha})$, which is in the
$i$th subband. The wave is partially reflected at the interface to
the nonmagnetic region. The wave function in the ferromagnetic
source electrode is then
\begin{eqnarray}\nonumber
\psi^{\text{f}}(x,y)=
\frac{\chi^{\text{f}}_i({\bf k}^{\text I}_i)}
{\sqrt{|v_i^{\text{f}} ({\bf k}^{\text I}_i)|}} 
e^{i k^{\text I}_i (x\cos{\alpha} + y\sin{\alpha})} \\
\label{psileft}
+ \sum_{n=1}^{4} r_{i,n} 
\frac{\chi^{\text{f}}_n({\bf k}^{\text R}_n)}
{\sqrt{|v_n^{\text{f}} ({\bf k}^{\text R}_n)|}} 
e^{-i k^{\text R}_n (x\cos{\alpha^{\text R}_n}
+ y\sin{\alpha^{\text R}_n})} \; ,
\end{eqnarray}
where $\chi_n^{\text{f}}$ are the eigenfunctions of $H_{\text{f}}$
in ${\bf k}$-space representation, the normalization factor
$v_n^{\text{f}} ({\bf k}_n)$ is the velocity expectation value
computed for the state vector $\chi^{\text{f}}_n ({\bf k}_n)$,
and $r_{i,n}$ are the reflection coefficients to be determined from
the matching. The wave vectors ${\bf k}^{\text R}_n=k^{\text R}_n\,
(\cos{\alpha^{\text R}_n},\sin{\alpha^{\text R}_n})$ of the
reflected waves are determined by the following three conditions:\\
a) the modulus $k^{\text R}_n$ is the solution of the implicit
equation $\epsilon^{\text{f}}_n (k^{\text R}_n)=E_{\text F}$, where
$\epsilon^{\text{f}}_n (k)$ is the dispersion relation;\\
b) the angles $\alpha^{\text R}_n$ are derived from the continuity
of the momentum parallel to the interface due to translational
invariance along that spatial direction;\\
c) among these solutions we allow those that satisfy   
$\{ \Re{[ v_n^{\text{f}} ({\bf k}^{\text R}_n) ]} > 0 
\; \& \; \Im{[ v_n^{\text{f}} ({\bf k}^{\text R}_n) ]} = 0 \}$
or $\{ \Im{[ v_n^{\text{f}} ({\bf k}^{\text R}_n) ]} > 0 \}$.
The two possibilities correspond to propagating and evanescent
modes, respectively. 

Similarly, the wave function in the nonmagnetic region is
\be
\label{psiright}
\psi^{\text{p}}(x,y)=\sum_{n=1}^{4} t_{i,n} 
\frac{\chi^{\text{p}}_n({\bf k}^{\text T}_n)}
{\sqrt{|v_n^{\text{p}} ({\bf k}^{\text T}_n) |}} 
e^{i k^{\text T}_n (x\cos{\alpha^{\text T}_n} +
y\sin{\alpha^{\text T}_n})} \; ,
\ee
where $t_{i,n}$ are the transmission coefficients, and all the
other quantities in Eq.~(\ref{psiright}) have the same definitions
given for the corresponding quantities in Eq.~(\ref{psileft}).
The eight coefficients $r_{i,n}$ and $t_{i,n}$ are determined
by the two conditions of continuity of the wave function 
and current conservation, Eq.~(\ref{condition}), given that
$\psi^{\text{f,p}}$ has four components in the total momentum
space.

We emphasize that it is important to include all modes even when
some of them are evanescent since transmission and reflection at
the interface are influenced by tunneling into classically
forbidden channels.

\section{Results of numerical simulations}
\label{s:numsim}

For our numerical simulations we take $J_{\text{pd}} =
0.06$~eV~nm$^3$, $N_{\text{Mn}} = 1$~nm$^{-3}$ and $S=5/2$. The
Rashba term is characterized by $\langle \beta E_{z} \rangle=
0.05$~eV~nm, and the Fermi energy is taken as $E_{\text F} =
0.08$~eV, which corresponds to a hole density of $4 \times
10^{16}$~m$^{-2}$. We assume that the leads' magnetization
direction is either perpendicular ($\theta = 0$) or within the
plane ($\theta= \pi/2)$ of the quantum well.

To clarify the underlying physics, we will approach the full
spin FET design step by step. First, we consider Rashba spin
precession for holes transmitted through a single interface between
a ferromagnetic and a nonmagnetic 2DHG for a fixed injection angle.
Then, we include a second interface with the magnetization of the
source and drain electrode being parallel, keeping the angle of
incidence for spin-polarized holes still fixed. As a result of the
Rashba effect, the total transmission will oscillate as a function
of the channel length in the nonmagnetic 2DHG. Finally, we take the
full 2D nature of the device into account by adding up the current
contributions for all injection angles.

\begin{figure}
\includegraphics[width=8cm]{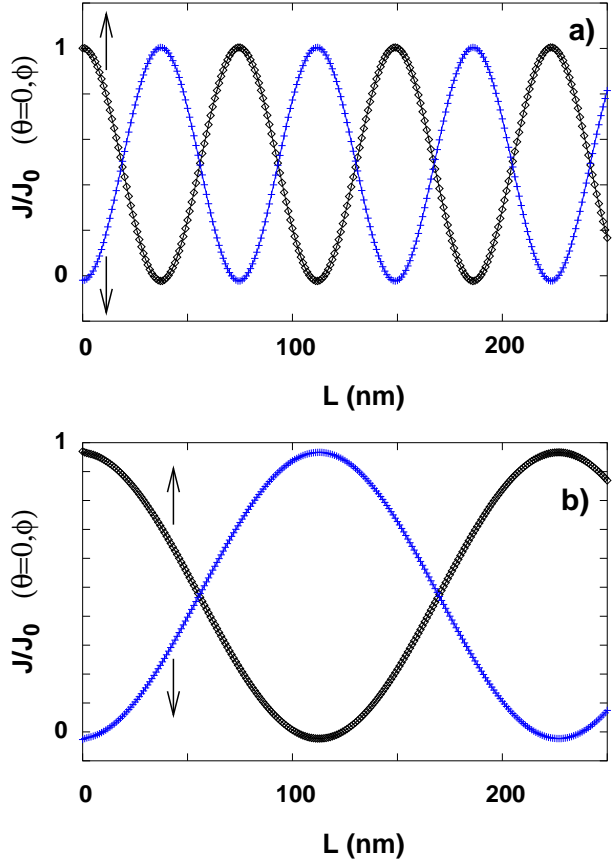}
\caption{Spin components of the current density in a nonmagnetic
2DHG plotted as a function of the distance $L$ from the interface
with a ferromagnetic 2DHG. Results are shown for a half--metallic
ferromagnetic contact having magnetization direction $\hat{{\bf n}}
=(0,0,1)$. The Fermi energy is equal to 0.09~eV in a) and to
$0.045 $~eV in panel b). It is apparent that the period of current
oscillations is proportional to $1/E_{\text F}$. For both spin
directions, the current is normalized to the incident hole flux.}
\label{f5}
\end{figure} 

Let us start by considering a single interface between a
ferromagnetic and a nonmagnetic 2DHG. In Fig.~\ref{f5} we show the
spin-up and spin-down currents as a function of the distance from
the interface in the case of magnetization perpendicular to the
plane ($\hat{{\bf n}}=(0,0,1)$) and perpendicular injection
($\alpha=0$) of spin-up current for two different values of
$E_{\text F}$. We find that both the spin-up and spin-down current
density oscillates with modulation length $L_{\text {so}}$,
indicating Rashba spin precession in the nonmagnetic region. With
increasing Fermi energy, the oscillation length decreases. This
result is in clear contrast to the case of the spin FET based on
electrons, where the spin precession length is independent of the
Fermi energy.\cite{spinfet1} Moreover, the modulation length
$L_{\text {so}}(\alpha)$ depends also on the injection angle
$\alpha$ (not shown in Fig.~\ref{f5}). For a realistic sample,
integration over all possible injection angles is required. We will
see later that, after integration, the overall modulation length is
given by that for perpendicular injection, $L_{\text {so}}(0)$.

\begin{figure}
\includegraphics[width=8cm]{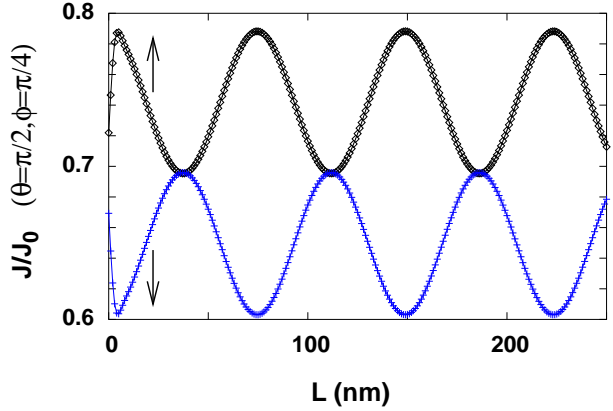}
\caption{Same as Fig.~\ref{f5} a) but for the case where the
magnetization direction in the ferromagnetic 2DHG (and, hence, the
quantization axis for spin components of the current) is equal to
$\hat{{\bf n}}=(1/\sqrt{2},1/\sqrt{2},0)$, i.e., lies in the 2DHG
plane. Note the diminished amplitude of current oscillations which
would disappear altogether for $\hat{{\bf n}}=(0,1,0)$.}
\label{f6}
\end{figure}

In Fig.~\ref{f6} we show results for the case of magnetization
direction in the ferromagnetic 2DHG being $\hat{{\bf n}}=(1/
\sqrt{2},1/\sqrt{2},0)$. For in--plane magnetization in the
half--metallic contact, the amplitude of the current modulation in
the nonmagnetic 2DHG depends on its azimuthal angle $\phi$. The
largest oscillation amplitude occurs for $\phi=0$, i.e., when the
magnetization direction is perpendicular to the interface. No
oscillations exist for $\phi=\pi/2$, because majority spins
injected into the nonmagnetic 2DHG are then eigenstates of
$H_{\text{rs}}$.

\begin{figure}[b]
\includegraphics[width=8cm]{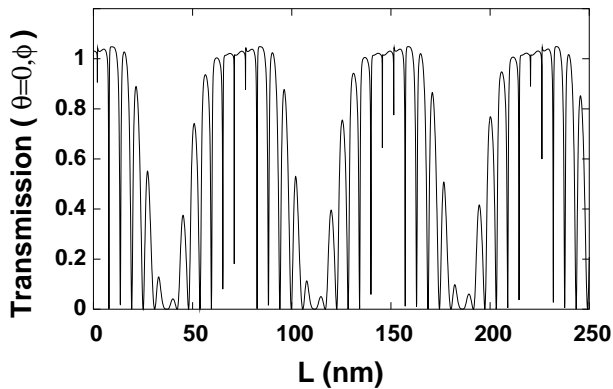}
\caption{The transmission probability of spin--up electrons for the
case of two interfaces separated by a distance $L$. The
high--frequency oscillations are due to resonances arising from
multiple reflections between the two interfaces. It turns out (see
below) that such features tend to be smeared out when the
transmission is averaged over the injection angle. Parameters are
the same as in Fig.~\ref{f5}.}
\label{f7}
\end{figure} 

We now turn to the simulation of the spin FET transistor,
consisting of a finite strip of nonmagnetic 2DHG with two
interfaces with ferromagnetic contacts, one at $x=0$ and the other
one at $x=L$. Now we have to apply the mode-matching procedure for
each interface. In Fig.~\ref{f7}, we plot the total transmission
through the entire device as a function of the width $L$ of the
nonmagnetic region. We find a modulation of the transmission with
modulation length $L_{\text {so}}$, which is due to Rashba spin
precession. This modulation is superimposed by fast oscillations
of the order of twice the Fermi wavelength $\lambda_{\text F}$
which are due to interference effects from multiple reflection
within the double-barrier structure.\cite{matsuyama} As we will see
below, these fast oscillations will be almost always smeared out
after integration over the injection angles $\alpha$, i.e., they
will not appear in real 2D devices. Only in the limit of very low
hole densities, remnants of these oscillations will be visible.

\begin{figure}
\includegraphics[width=8cm]{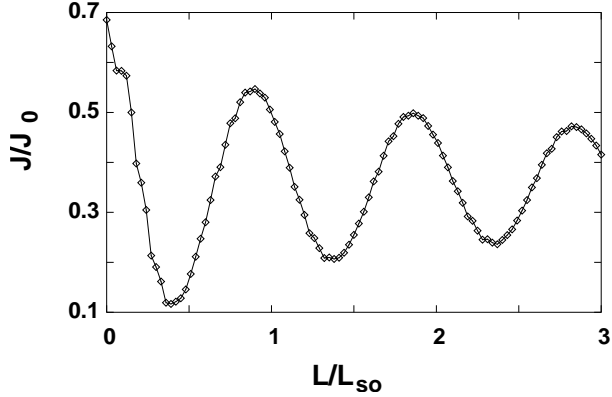}
\caption{The total current density as a function of the quantity
$L/L_{\text {so}}$ in the 2D system treatment, where $L$ is the
channel length and $L_{\text {so}}$ is the total modulation length.
The magnetization in the contacts is perpendicular to the plane of
the 2DHG.}
\label{f8}
\end{figure}

Finally, we take into account the full 2D nature of the device,
i.e., we add up the current contributions for all injection angles.
We assume an isotropic angular distribution of injected holes
since all our simulations are performed in the linear response
regime. As a result, the transmitted current density $J/J_0$ is
given by the formula
\be
\label{integration}
J/J_0 = \sum_{n,m} \int_{-\pi/2}^{\pi/2} T_{nm}
(\alpha) \cos{\alpha} \, d \alpha \;,
\ee
where $T_{nm}(\alpha)$ is the transmission probability from channel
$n$ to $m$ for holes injected at an angle $\alpha$. Since the
modulation length $L_{\text {so}}(\alpha)$ of the transmission is
$\alpha$-dependent, one might expect that the integration washes
out the effects of the spin precession. It turns out, however, that
oscillations are still visible, although damped. The modulation
length of the resulting oscillations coincides with that for
perpendicular incidence, $L_{\text {so}}(0)$. In Fig.~\ref{f8} we
show the result of the integration operation corresponding to
Eq.~(\ref{integration}) for magnetization direction in the contact
2DHGs being $\hat{\bf n}=(0,0,1)$.

\section{Analytical results for transport}
\label{s3}

In this section we show that most features of the numerical results
presented in the previous section can be understood within a
simplified analytical model. In particular, we derive analytic
expressions for the precession length $L_{\text {so}}(\alpha)$ and
the total current density. A similar model has been already
proposed in Ref.~\onlinecite{pala}, where it was used to discuss
universal features of hole and electron spin precession. The
approximate formulae derived in this section clearly show how the 
precession length depends on system parameters and, therefore,
allow for a deeper understanding of the underlying physics than
looking at the purely numerical results presented in the previous
Section can provide.

The model is developed following some approximations that are
justified {\it a priori} by physical considerations and {\it a posteriori}
by the comparison between analytical and numerical results. First,
we make use of the fact that for typical parameters only the lowest
heavy-hole subband is occupied. As an approximation we can,
therefore, omit all non-conducting subbands from our model.
Furthermore, we assume perfect transmission at the interfaces,
i.e., we neglect reflection. All Hamiltonians are now represented
as $2\times 2$ matrices, using the basis (\ref{basis}) restricted
to vectors $| 1 \rangle$ and $| 4 \rangle$. The off-diagonal matrix
elements are obtained from perturbation theory for the degenerate
case.\cite{bir} The nonmagnetic 2DHG region is then described by
\be
\label{newhso}
H_{\text{p}}= \frac{\hbar^2 k^2}{2m_{\text{hh} \parallel}}
\left( \begin{array}{cc}
1 & 0\\
0& 1
\end{array} \right)
+
i \langle \beta_{\text h} E_{z} \rangle k^3 \left(
\begin{array}{cc}
0 & e^{-i3\alpha}\\
e^{i3\alpha} & 0
\end{array} \right) \; ,
\ee
where $\beta_{\text h}$ is proportional to the spin-orbit coupling 
of holes and is different from the $\beta$ defined in
Eq.~(\ref{hrs}).

The corresponding eigenenergies are
\be
\label{eps}
\epsilon_{1,2}(k)=\frac{\hbar^2}{2m_{\text{hh} \parallel}} k^2 \pm 
\langle \beta_{\text h} E_{z} \rangle k^3
\ee
with eigenvectors
\be
\chi_{1,2}=\frac{1}{\sqrt{2}} \left( \begin{array}{c} 1 \\ 
\mp i e^{i 3 \alpha} \end{array} \right) \; .
\ee
The spin splitting of the eigenvalues (\ref{eps}), together with
the conservation of the wave vector parallel to the interface,
implies the presence of a double refraction
phenomenon\cite{matsuyama,biref} where a hole wave incident on the
interface from the ferromagnet gives rise to two transmitted waves
in the nonmagnetic 2DHG having slightly different wave vectors.
Their magnitudes $k_{1,2}$ are obtained from the implicit equation
$\epsilon_{1,2}(k) = E_{\text F}$. Typically, spin-orbit coupling
can be treated as a perturbation, which means that we can linearize
the expression of $k_{1,2}$ in the spin-orbit coupling strength,
and arrive at $k_{1,2}=k_{0} \mp \Delta k/2$. Here $k_{0}$ is the
Fermi wave vector in the absence of spin-orbit coupling, and
\be
\Delta k = \left( \frac{2 m_{\text{hh} \parallel}}{\hbar^2}
\right)^2\langle \beta_{\text h} E_{z} \rangle E_{\text F} \, ,
\ee
which explicitly depends on the Fermi energy. The corresponding
angles of the transmitted waves' propagation direction with the
interface normal are found, again in the limit of weak SO coupling,
to be 
\be
\alpha_{1,2}=\alpha_0 \pm (\Delta k/2k_{0}) \tan{\alpha_0} \, ,
\ee 
where $\alpha_0$ is defined by
\be
\label{a0}
k_{\text{F}} \sin\alpha=k_{0} \sin \alpha_0 \; . 
\ee
Hence, the transmitted hole is described by the wave function
\be
c_{1}  \chi_{1} e^{i k_{1} (x \cos\alpha_1  +y \sin\alpha_1)}
+c_{2} \chi_{2} e^{i k_{2} (x \cos\alpha_{2}  +y \sin
\alpha_{2})}\, .
\ee 
By assuming a perfectly transparent interface, we can compute the
coefficients $c_{1,2}$ simply by matching the wave functions in the
ferromagnet and in the nonmagnetic semiconductor. At the interface
at $x=L$ to the second ferromagnet, for the case of its
magnetization pointing in positive $z$ direction, only the
$|+\rangle$ component will be transmitted. Hence, the outgoing
state in the right ferromagnet reads $e^{i k_{\text{F}} (x \cos
\alpha  +y \sin\alpha)} \cos[\Delta k L/ (2\cos \alpha_0)] |+
\rangle$. As a result, the transmission probability is
\begin{equation}
\label{tperp}
T_{0,\phi}(\alpha)=\cos^2\left[\frac{\gamma}{\cos \alpha_0}
\right] \; , 
\end{equation} 
where we have used the relation $\Delta k\, L/2=\gamma$, and the
dependence on $\alpha$ is through $\alpha_0$ via Eq.~(\ref{a0}).
In a similar way we can obtain the transmission probabilities 
for arbitrary magnetization direction in the ferromagnetic 2DHGs.
The transmission probability for in-plane magnetization reads 
\begin{equation}
\label{tinplane}
T_{\pi/2,\phi}(\alpha)=\cos^2\left[\frac{\gamma}{\cos \alpha_0}
\right] + \sin^2\left[3\alpha_0-\phi\right]\sin^2
\left[\frac{\gamma}{\cos \alpha_0}\right] \; .  
\end{equation} 
Finally, we can write the transmission for arbitrary magnetization 
direction as
\begin{equation}
\label{tgeneral}
T_{\theta,\phi}(\alpha)=\cos^2\theta\, T_{0,\phi}+\sin^2\theta\,
T_{\pi/2,\phi} \; .
\end{equation} 
Equations.~(\ref{tperp}-\ref{tgeneral}) cease to be valid once
one of the transmitted states in the nonmagnetic 2DHG becomes
evanescent, i.e., is totally reflected. This condition defines 
critical angles $\alpha_{\text{c},\{1,2\}}$, that in the limit of
weak SO coupling read $\alpha_{\text{c},\{1,2\}}=k_{0}/k_{\text{F}}
\mp \frac{1}{2} \Delta k/k_{\text{F}} \approx k_{0}/k_{\text{F}}=
\alpha_c$. We note that very similar formulae for the transmission
can be obtained for electrons.\cite{pala}

From Eqs.~(\ref{tperp}) and (\ref{tinplane}) we find the precession
length 
\be
L_{\text {so}} (\alpha)=\frac{2\pi} {
\langle \beta_{\text h} E_{z} \rangle} 
\left(\frac{\hbar^2}{2m_{\text{hh} \parallel}}\right)^2 \frac{
\cos{\alpha_0}}{E_{\text F}}
\ee
which depends on both the injection angle $\alpha$ and the Fermi
energy $E_{\text F}$. The physical reason for the latter dependence
is the cubic spin splitting of the heavy-hole subband. For a 2D
device under consideration in this paper, one has to integrate over
all injection angles, see Eq.~(\ref{integration}).
The final result reads then
\begin{eqnarray}
\nonumber
& &J/J_0= \Big\{ \cos^2 \theta F(\gamma)+\sin^2 \theta \\
\label{current}
& & \times \Big[
\sin^2 \phi+ F(\gamma)\cos^2 \phi+ 
 G(\gamma)\cos (2 \phi)
\Big]\Big\} \;,  
\end{eqnarray}
where $J_0$ is the injected current density, and the functions
$F(\gamma)$ and $G(\gamma)$ are defined as
\begin{eqnarray}
\label{fint}
&F(\gamma)=\frac{1}{2}\int_{-\frac{\pi}{2}}^{\frac{\pi}{2}}
\cos{\alpha} \cos^2{\left(\frac{\gamma}{\cos{\alpha}}\right)}\,
d\alpha\, , &\\
\label{gint}
&G(\gamma)= 
\frac{1}{2}\int_{-\frac{\pi}{2}}^{\frac{\pi}{2}} \cos{\alpha} 
\sin^2{(3\alpha)}
\sin^2{\left(\frac{\gamma}{\cos{\alpha}}\right)}\,d\alpha&  \; .
\end{eqnarray}
A good analytical approximation for $F(\gamma)$ and $G(\gamma)$ is
given in Ref.~\onlinecite{pala}. The result Eq.~(\ref{current})
describes damped oscillations of the current density as a function
of the length of the nonmagnetic part, where the modulation length
is given, in agreement with our numerical findings, by the
precession length for perpendicular injection, $L_{\text {so}}
(\alpha=0)$. 

\section{Gate-voltage manipulation of the current: Capacitance
model}
\label{s4}

In this section we consider the response of the hole spin FET to 
external gate voltages and point out important differences in its
behavior compared to the electron version. With standard
densities of about $10^{16}$~m$^{-2}$, only the lowest spin--split
2D HH subbands are occupied. Their $k^3$ spin splitting leads to an
inversely linear dependence on the Fermi energy $E_{\text F}$
for the hole precession length $L^{\text h}_{\text{so}}$. Hence,
variation of gate voltages will modify $L^{\text h}_{\text{so}}$
by changing, at the same time, the asymmetry of the hole
confinement in the 2DHG and the Fermi energy. We analyze how the
device performance changes when, instead of only a top gate
voltage, both top gate and back gate voltages are applied.

\begin{figure}
\includegraphics[width=8cm]{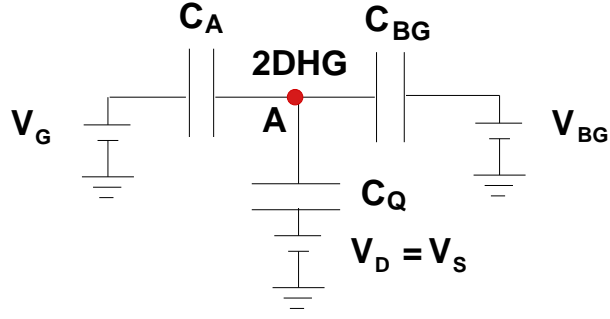}
\caption{Linearized capacitance model for the influence of top and
back--gate voltages on the spin FET. The 2DHG is located at point
A, $V_{\text G}$ is the voltage applied at the top gate,
$V_{\text{BG}}$ that of the back gate. The voltages at the drain
($V_{\text{D}}$) and source ($V_{\text{S}}$) contacts are assumed
to be equal since we consider the linear response regime.}
\label{f9}
\end{figure} 

We model the effect of the top and back-gate voltages 
through a linearized capacitance model as shown in Fig.~\ref{f9}. The
capacitance per unit area between the top gate and the point A,
where the 2DHG is located, is $C_{\text A}=\epsilon_0 
\epsilon_{\text{GaAs}}/d_{\text A} \sim 2 \times 10^{-3}$~F/m$^2$ 
for an effective distance $d_{\text A}=50$~nm between top gate and 2DHG, 
while the capacitance associated to the variation of the back-gate
voltage is $C_{\text{BG}}=C_{\text A}/2$ for a distance of 100~nm
from the 2DHG. The capacitance between the 2DHG and the source and
the drain is well approximated by the quantum capacitance 
$C_{\text Q}=m_{\text{hh} \parallel}e^2/\pi\hbar^2 \sim 8 \times
10^{-2}$~F/m$^2$ and is due to the finite density of states in the
2DHG.\cite{luryi} The variation of the voltage at the point A 
and of the electric field $E_z$ read then
\begin{eqnarray}\nonumber
dV_{\text A} & = & 
\frac{C_{\text A}}{C_{\text A} + C_{\text Q}+C_{\text BG}}
dV_{\text G} + \frac{C_{\text{BG}}}{C_{\text A}+C_{\text Q}+
C_{\text{BG}}} dV_{\text{BG}} \\ \nonumber 
dE_{z} & = & 
-\frac{C_{\text Q}+C_{\text{BG}}}{\epsilon_0
\epsilon_{\text{GaAs}}} (dV_{\text A}-dV_{\text{BG}})\\ \nonumber
 & = & -\frac{ (C_{\text Q}+C_{\text{BG}})   
[C_{\text A} dV_{\text G}-(C_{\text A}+C_{\text Q})dV_{\text{BG}}]}
{\epsilon_0 \epsilon_{\text{GaAs}}
(C_{\text A}+C_{\text Q}+C_{\text{BG}})}
 \; ,
\end{eqnarray}
while the variation of the Fermi energy reads
\be
\label{def}
dE_{\text F}=-e \, dV_{\text A}=\frac{-e\,(C_{\text A} dV_{\text G}
+C_{\text{BG}} dV_{\text{BG}})}{C_{\text Q}+C_{\text{BG}}+
C_{\text A}} \; .
\ee
 
\begin{figure}[b]
\includegraphics[width=8cm]{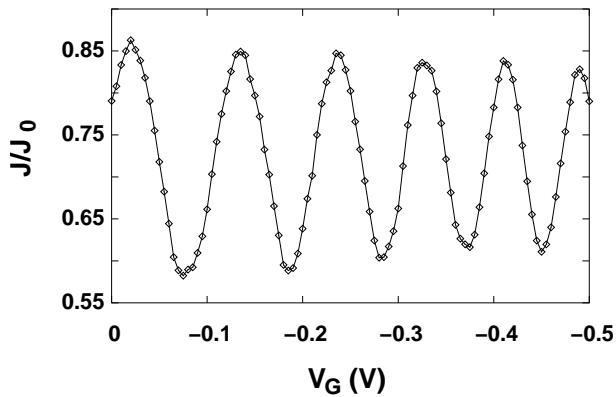}
\caption{The modulation of the current density through the spin FET
when the top-gate voltage is varied. The magnetization direction in
the ferromagnetic 2DHGs is given by $\hat{{\bf n}}=(0,0,1)$. The
Fermi energy for $V_{\text G}=0$ is $E_{\text F}=0.09$~eV, the
Rashba coupling term $\langle \beta E_{z} \rangle = 0.05$~eV~nm.}
\label{f10}
\end{figure} 

\begin{figure}[h!]
\includegraphics[width=8cm]{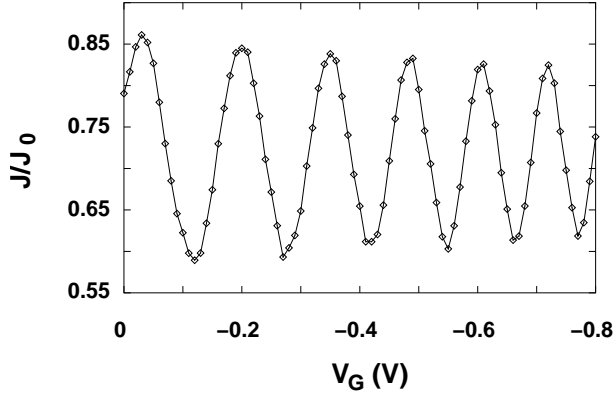}
\caption{The modulation of the current density in the spin FET when
the top-gate voltage and the bulk-gate voltage are simultaneously
varied in a way such that $E_{\text F}$ is unchanged. The contact
magnetization points along $\hat{{\bf n}}=(0,0,1)$.}
\label{f11}
\end{figure} 

It is clear that variation of only $V_{\text G}$ (i.e., keeping
$dV_{\text{BG}}=0$) simultaneously changes $E_{z}$ and
$E_{\text F}$. In order to leave the Fermi level pinned, we have to
manipulate both top and back-gate voltages such that
$dV_{\text{BG}}=-(C_{\text A}/C_{\text{BG}})\,dV_{\text G}$.
Results of our simulation for the case that only the top-gate
voltage is varied ($dV_{\text{BG}}=0$) is shown in Fig.~\ref{f10}.
Here the current density obtained from Eq.~(\ref{integration}) is
plotted as function of the gate voltage $V_{\text G}$ for the case
when the magnetization direction in the contacts is perpendicular
to the 2DHG. Spin--precession--induced current oscillations are
clearly visible. The oscillation period actually varies with
changing $V_{\text G}$ due to the induced variation of
$E_{\text F}$, as mentioned in Sec.~\ref{s2}. Note that, for the
parameters of Fig.~\ref{f10}, the effect of the gate voltage 
on the position of subband bottoms is negligible since the
electric field is changed only by a few percent from its initial
value $E_z \sim e p/\epsilon_0 \epsilon_{\text{GaAs}}$. (Here $p$
is the hole density.) For comparison, we show results for the case
where both top and back-gate voltages are varied simultaneously
such that the hole density in the 2DHG remains unchanged in
Fig.~\ref{f11}. Here the precession length changes only due to the
gate--voltage--induced variation of the structural inversion
asymmetry, measured here by the electric field $E_{z}$. Current
oscillations have then a larger period as function of $V_{\text G}$
than in the case where $V_{\text{BG}}$ is kept constant. In both
cases, however, a clear modulation of the current as a function of
gate voltage $V_{\text G}$ is obtained. These are slightly damped
due to the superposition of current amplitudes for all possible
angles of incidence.

\section{CONCLUSIONS}

We have performed careful numerical and analytical studies of
transport through a p-type all--semiconductor spin FET. 
The design of such a device would overcome problems associated 
with the fabrication of hybrid devices involving metal--semiconductor
contacts. Despite the more complicated nature of spin splitting in
2D valence--band states, clear current modulation as a function of
device parameters such as the width of the nonmagnetic region are
observed. Using a phenomenological model for the action of external
gate voltages, we have shown the possibility of current
manipulation as was envisioned in the original electron spin FET
proposal by Datta and Das.\cite{spinfet1} Our numerical simulations
that were performed for realistic sample parameters, as well as
analytical formulae reported in this work, should serve as a
useful basis for experimental realization and functional
optimization of a p--type spin FET.

\begin{acknowledgments}

This work was supported by Deutsche Forschungsgemeinschaft via the
Center for Functional Nanostructures at the University of Karlsruhe,
by the Emmy-Noether program,
by the IST NanoTCAD project (EC contract IST-1999-10828) 
and by EU Research Training Network
RTN2-2001-00440.

\end{acknowledgments}


 
\end{document}